\documentclass{emulateapj}
\usepackage{times}

\def\F{{\it Fermi}-LAT }
\def\deg{\hbox{$^\circ$}}
\def\arcsec{\hbox{$^{\prime\prime}$}}

\shorttitle{X-ray and optical observations of 1FGL J1311.7-3429}  
\shortauthors{J.Kataoka et al.}

\begin{document}

\title{Toward Identifying the Unassociated Gamma-ray Source 1FGL J1311.7-3429
with X-ray and Optical Observations}

\author{
J.~Kataoka\altaffilmark{1,2}, 
Y.~Yatsu\altaffilmark{3}, 
N.~Kawai\altaffilmark{3}, 
Y.~Urata\altaffilmark{4}, 
C.~C.~Cheung\altaffilmark{5}, 
Y.~Takahashi\altaffilmark{1}, 
K.~Maeda\altaffilmark{1}, 
T.~Totani\altaffilmark{6}, 
R.~Makiya\altaffilmark{6},
H.~Hanayama\altaffilmark{7},
T.~Miyaji\altaffilmark{7}, 
A.~Tsai\altaffilmark{4}
}

\altaffiltext{1}{Research Institute for Science and Engineering, Waseda University, 3-4-1, Okubo, Shinjuku, Tokyo 169-8555, Japan}
\altaffiltext{2}{email: kataoka.jun@waseda.jp}
\altaffiltext{3}{Tokyo Institute of Technology, 2-12-1, Ohokayama, Meguro, Tokyo 152-8551, Japan}
\altaffiltext{4}{Institute of Astronomy, National Central University,
Chung-Li 32054, Taiwan}
\altaffiltext{5}{National Research Council Research Associate, National Academy of Sciences, Washington, DC 20001, resident at Naval Research Laboratory, Washington, DC 20375, USA}
\altaffiltext{6}{Department of Astronomy, Kyoto University,
Kitashirakawa, Sakyo-ku, Kyoto 606-8502, Japan}
\altaffiltext{7}{Ishigakijima Astronomical Observatory, National Astronomical Observatory of Japan, 1024-1 Arakawa, Ishigaki, Okinawa, 907-0024, Japan}

\begin{abstract}
We present deep optical and X-ray follow-up observations of the bright 
unassociated $Fermi$-LAT gamma-ray source 1FGL J1311.7-3429. 
The source was already known as an unidentified 
EGRET source (3EG J1314-3431, EGR J1314-3417), hence its nature 
has remained uncertain for the past two decades.  
For the putative counterpart, we detected a quasi-sinusoidal optical modulation
of $\Delta m$ $\sim$ 2 mag with a period of $\simeq$1.5 hr 
in the $Rc$, $r'$ and $g'$ bands. Moreover, we found that 
the amplitude of the modulation and peak intensity 
changed by $\gtrsim$ 1 mag and  $\sim$0.5 mag respectively,  
over our total six nights of observations from 2012 March and May.
Combined with $Swift$ UVOT data, the optical-UV 
spectrum is consistent with a blackbody temperature, $kT$ $\simeq$ 1 eV, 
and the emission volume radius $R_{bb}$ $\simeq$ 
1.5$\times$10$^4$ $d_{\rm kpc}$ km ($d_{\rm kpc}$ is 
the distance to the source in units of 1 kpc).  
In contrast, deep $Suzaku$ observations conducted in 2009 and 2011 
revealed strong X-ray flares with a lightcurve characterized 
with a power spectrum density of $P(f)$ $\propto$ $f^{-2.0\pm0.4}$, but 
the folded X-ray light curves suggest an orbital modulation also in X-rays.
Together with the non-detection of a radio counterpart, and
significant curved spectrum and non-detection of variability in
gamma-rays, the source may be the second ``radio-quiet''
gamma-ray emitting milli-second pulsar candidate after 1FGL J2339.7-0531,
although the origin of flaring X-ray and optical
variability remains an open question.
\end{abstract}

\keywords{gamma rays: stars --- pulsars: general --- X-rays: general}

\section{Introduction}

The Large Area Telescope \citep[LAT;][]{atw09} 
onboard the $Fermi$ Gamma-ray Space 
Telescope is a successor to EGRET onboard the Compton Gamma-ray
Observatory \citep{har99}, with much improved sensitivity, 
resolution, and energy 
range. The second \F catalog, based on the first 24 months of all-sky survey data 
\citep[2FGL;][]{2FGL},  
provides source location, flux and spectral information, 
as well as light curves on month time bins for 1873 $\gamma$-ray sources 
detected and characterized in the 100 MeV to 100 GeV range. Thanks to their 
small localization error circles (or ellipses) with typical 95$\%$ confidence radii, $r_{95}$ 
$\simeq 0.1\deg-0.2\deg$, for relatively bright sources, 69 $\%$ 
of the 2FGL sources are reliably
associated or firmly identified with counterparts of known or likely $\gamma$-ray producing 
sources. In particular, more than 1,000 sources are proposed to be
associated with AGN (of mainly the blazar class) and 87 sources with 
pulsars (PSRs), 
including 21 millisecond pulsars (MSPs) which are a
new category of $\gamma$-ray sources discovered with \F
\citep{2FGL, MSP}. 
Other sources, albeit of a relative minority compared to AGNs and PSRs, 
also constitute important categories of new GeV sources like 
supernova remnants \citep[SNRs;][]{W51, Cas-A, W44, IC443},  
low-mass/high-mass binaries \citep{LS5039, LSI, CygX-3}, 
pulsar wind nebula \citep{Vela, 1509}, one nova \citep{V407}, 
normal and starburst galaxies \citep{LMC, M82}, and the giant lobes 
of a radio galaxy \citep{cena-lobe}.

Despite such great advances in the identification of \F sources, 
575 (31$\%$) sources in the 2FGL catalog still remain unassociated. 
Note that a substantial fraction of the unassociated sources (51$\%$) 
have at least one analysis flag  due to various 
issues, while only 14$\%$ of the associated sources have been 
flagged  \citep{2FGL}. This may suggest some of unassociated sources 
are spurious due to complexity/difficulty of being situated in a 
crowded region near the Galactic plane. Nevertheless, many of them 
are bright enough to be listed in the one year \F catalog
\citep[1FGL;][]{1FGL} and some of them are even listed in the bright 
source list based on the first 3 months of data \citep[0FGL;][]{0FGL}.
By comparing the distribution of associated and unassociated sources 
in the sky, a number of interesting features in the map were reported 
\citep{2FGL}. 
For example, (1) the number of unassociated sources decreases 
with increasing Galactic latitude, (2) the number of unassociated
sources increases sharply below Galactic latitudes, $|b|$ $<$ 10$^{\circ}$, 
and (3) the fraction of sources with curved gamma-ray spectra
among the unassociated sources is greater (28$\%$) than the fraction 
of curved spectra sources among the associated sources (16$\%$). Further 
extensive studies based on a statistical approach in an effort to
correlate their gamma-ray properties with the AGN and  PSR populations 
was presented for 1FGL unassociated sources \citep{UnIDstat}.

In this context, 1FGL J1311.7-3429 (or 2FGL name, 2FGL J1311.7-3429) 
is a $classical$ unassociated gamma-ray source situated at 
high Galactic latitude ($l$ = 307.6859\deg, $b$ = 28.1951\deg), and was first 
discovered by EGRET about 20 years ago as 
3EG J1314-3431 \citep{har99} or EGR J1314-3417 \citep{cas08}.
The source was also reported by \F in the 0FGL list with a gamma-ray 
flux $F_{\rm 0.1-20 GeV}$ = (11.7 $\pm$1.1) $\times$ 10$^{-8}$ photons
cm$^{-2}$ s$^{-1}$, which is marginally consistent with the gamma-ray 
flux determined by EGRET, $F_{\rm 0.1-20 GeV}$ = (18.7 $\pm$3.1) $\times$ 
10$^{-8}$ photons cm$^{-2}$ s$^{-1}$, within the 2$\sigma$ level.
In the 2FGL catalog, the detected significance of 1FGL J1311.7-3429 
is 43.1 $\sigma$, which is one of the brightest sources with an \textsc{unid} 
flag.  Based on the one-month binned $>100$ MeV gamma-ray 
light curve of 1FGL J1311.7-3429 over two years of data as 
published in the 2FGL
catalog\footnote{http://heasarc.gsfc.nasa.gov/FTP/fermi/data/lat/catalogs/source/lightcurves/\\
2FGL$\_$J1311d7m3429$\_$lc.png},
no statistically significant variability was observed
(\textsc{Variability\_index} = 19.09; \cite{2FGL}).
The gamma-ray spectrum is significantly 
curved with curved significance \textsc{Signif\_curve} of 6.33 \citep{2FGL}.
\footnote{As detailed in \cite{2FGL}, the \textsc{Variability\_index}
is an indicator of the flux being constant across the full 2-year
period. A value of \textsc{Variability\_index} $>$ 41.6 is 
used to identify variable sources at a 99$\%$ confidence level.
Similarly, the \textsc{Signif\_curve} parameter is an indicator of the spectrum
being curved, by comparing the likelihood values calculated for a LogParabola 
and a single power-law function. \textsc{Signif\_curve} is distributed 
as $\chi^2$ with one degree of freedom, and \textsc{Signif\_curve} $>$ 16
corresponds to 4 $\sigma$ significance of curvature.}  

The first X-ray follow-up observation of 1FGL J1311.7-3429 was conducted 
as a part of $Suzaku$ X-ray observations of 11 unidentified \F 
objects at high Galactic latitude, $| b |$ $>$ 10$^{\circ}$ 
\citep{mae11,tak12}. 
The X-ray source associated with 1FGL J1311.7-3429 showed a very 
rapid X-ray flare with the count rate
changing by a factor of 10. Subsequent $Chandra$ ACIS-I 
(2010 March 21 for a 19.87 ks  exposure, obsID 11790)
and $Swift$ XRT observations 
(2009 February 27 for a 3.34 ks exposure, obsID 31358; see also Table 1)
confirmed that the brightest X-ray source within the \F 
error ellipse is the most credible counterpart and that the X-ray 
source is also variable on month-to-year timescales. 
The unabsorbed X-ray flux observed with $Chandra$ was 1.03$\times$
10$^{-13}$ erg cm$^{-2}$ s$^{-1}$ in the 0.5$-$8 keV band, with a differential 
photon spectral index, $\Gamma$ = 1.26$^{+0.38}_{-0.37}$ \citep{che12}.

Motivated by the initial X-ray results, we conducted further 
deep observations of 1FGL
J1311.7-3429 with  $Suzaku$, together with deep optical 
observations using a 105 cm Ritchey-Chr\'etien 
telescope ($g'$, $Rc$, and $Ic$ bands) at the Ishigakijima Astronomical Observatory (IAO) in Japan, 
as well as the 1m telescope ($r'$ band) at Lulin Observatory in Taiwan.  
In Section 2, we describe the details of the $Suzaku$ observations and 
optical observations and data reduction procedures. 
Very recently, \cite{rom12} reported quasi-sinusoidal optical 
modulation of this source with a 1.56 hr (5626 s) period, 
suggesting that the source is another black-widow-type MSP 
like that recently discovered for 1FGL J2339.7-0531 \citep{rom11,kon12}.
Our paper confirms some of those optical findings for 1FGL
J1311.7-3429, plus provides results from multiple epoch 
optical monitoring between 2012
March and May and completely 
new X-ray data based on a long $Suzaku$ observation 
conducted in 2011 together with our previously published archival 2009
data. The results of these observations are given 
in Section 3. 
Based on our new observational data in
optical and X-ray, and various observed gamma-ray parameters compiled in  
the 2FGL source catalog, we support that 1FGL J1311.7-3429 could be a 
``radio-quiet'' gamma-ray emitting MSP candidate like 
1FGL J2339.7-0531. 
The variable optical/X-ray source is posited 
as the counterpart to the gamma-ray source and throughout, 
we refer to it simply as 1FGL J1311.7-3429.

\begin{center}
\begin{deluxetable*}{lcccc}{ht}
%\tabletypesize{\scriptsize}
%\rotate
\tablecaption{Observation log of 1FGL J1311.7-3429 analyzed in this paper}
\tablewidth{0pt}
\tablehead{
\colhead{Obs Start (UT)} &  \colhead{Obs End (UT)} & \colhead{Observatory} &
 \colhead{Band} & \colhead{Exposure} 
}
\startdata
2009-08-04 04:56:35 & 2009-08-05:07:18:14  & $Suzaku$ XIS & X-ray & 33.0 ks \\
2011-08-01 16:48:20 & 2011-08-03:17:40:15  & $Suzaku$ XIS & X-ray & 65.2 ks \\
2009-02-27 18:41:59 & 2009-02-27:22:04:57  & $Swift$ UVOT & UV & 276/276/276/552/742/1106$^a$ \\
2012-03-24 17:45:37 & 2012-03-24 19:38:35  & LOT 1m  & $r'$ & 5 min $\times$ 21 \\
2012-03-25 17:48:26 & 2012-03-25 19:26:33  & LOT 1m  & $r'$ & 5 min $\times$ 12 \\
2012-03-26 17:25:14 & 2012-03-26 18:29:59 & LOT 1m  & $r'$ & 5 min $\times$ 12 \\
2012-04-12 16:35:08 & 2012-04-12 18:24:30 & LOT 1m  & $r'$ & 5 min $\times$ 21 \\
2012-05-24 13:00:49 & 2012-05-24 15:34:44  & LOT 1m  & $r'$ & 5 min $\times$ 29 \\
2012-05-25 12:31:11 & 2012-05-25 15:41:11 & IAO 105cm & $g'$, $Rc$, $Ic$ & 20 min $\times$ 7  \\

\tableline
\enddata 
\tablecomments{$^a$: $Swift$ UVOT exposures for the $v/b/u/uvw1/uvm2/uvw2$
 bands in seconds.}
\end{deluxetable*}
\end{center}

\section{Observations and Data Reduction}
\subsection{Optical/UV}

As discussed in \cite{che12}, the brightest 
X-ray source within the \F error ellipse, CXOU J131145.71-343030.5, 
is the most credible counterpart and exactly the one detected in our  
previous (AO4; below) $Suzaku$ observation \citep{mae11}. The source was also detected with 
the $Swift$ XRT (observation ID 31358). Within the $Chandra$ error 
circle (0.6\arcsec\ at 90$\%$ level), there is a $R$ = 17.9 mag, 
$B$ = 20.5 mag star in the USNO A-2.0 catalog \citep{mon03a}, 
that is 0.56\arcsec\ apart from CXOU J131145.71-343030.5. 
The same star is also listed as $R$ = 18.8 mag, $B$ = 21.0 mag in 
USNO-B1.0 catalog, suggesting a hint of temporal variability 
\citep[a typical uncertainty of these measurements is $\simeq$ 0.3
mag;][]{mon03a}. The same optical source is also seen 
in the DSS (Digital Sky Survey), but its optical magnitude is unclear. 
Moreover, $Swift$ UVOT observations taken simultaneously with the XRT 
detected the source. Our analysis using archival $Swift$ UVOT
data indicate: $v$ $>$ 19.51 (upper limit only), 
$b$ = 20.10$\pm$0.30, $u$ = 20.77$\pm$0.27, $uvw1$ = 
21.70$\pm$0.31, $uvm2$ = 21.58$\pm$0.24, and 
$uvw2$ = 22.05$\pm$0.22.

\begin{figure}
\begin{center}
\includegraphics[angle=0,scale=0.31]{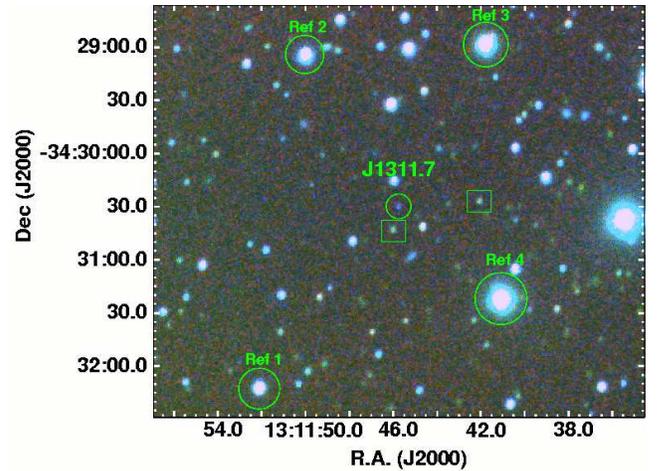}
\caption{Tri-color optical image of 1FGL J1311.7-3429 constructed
 from  data taken in the $g'$, $Rc$, and $Ic$-bands by the 
IAO 1.05 m telescope (Table~1). The four reference stars 
used in the analysis of the IAO data, are denoted as Ref 1-4, 
while two refrence stars used in the analysis of the LOT data 
are shown as $boxes$.}
\end{center}
\end{figure}

\begin{figure}
\begin{center}
\includegraphics[angle=0,scale=0.45]{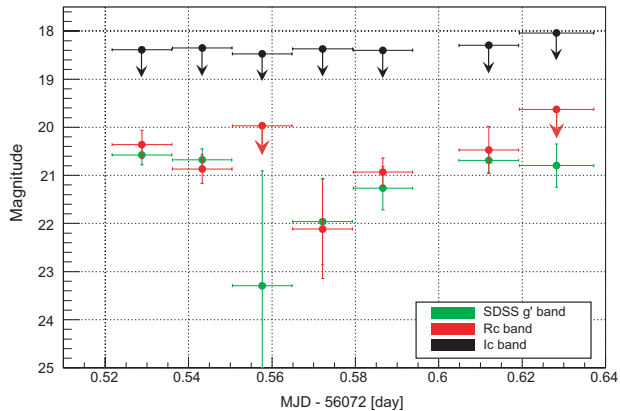}
\caption{Multi-band optical light curves of 1FGL J1311.7-3429 
observed with the IAO 1.05 m telescope on May 25 ($g'$, $Rc$ and $Ic$ bands). 
Data are binned by 1200 s (300 s $\times$ 4 frames) for the 
IAO data. The source is not detected in the $Ic$-band and 3 $\sigma$ upper 
limits are provided. }
%($bottom$) Folded optical lightcurve with suggestive orbital period of
 %5626.00 sec (Romani et al. 2012) taken in I, R, and G-bands (actually
 %G-bands is not detected and provides only 3 $\sigma$ upper limit).
\end{center}
\end{figure}

\begin{figure}
\begin{center}
\includegraphics[angle=0,scale=0.47]{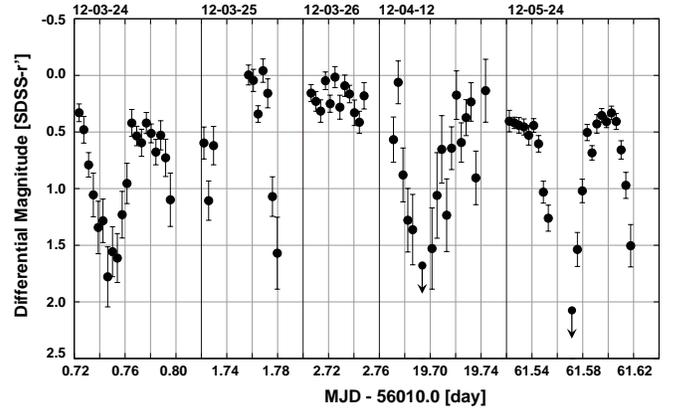}
\includegraphics[angle=0,scale=0.45]{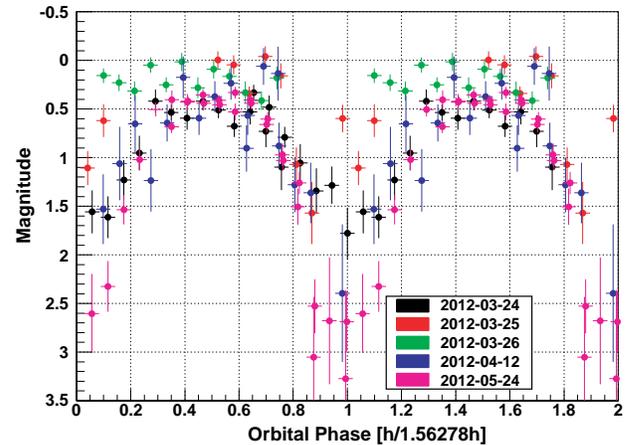}
\caption{ ($top$) Temporal variations in the differential  $r'$-band 
magnitudes derived by comparison with a reference star 
in the field, observed with the Lulin 1m telescope from 2012 
March to May. The $r'$-band magnitude of 
reference star remains constant within $\Delta$$m$ = 0.068 mag. 
Note that the modulation profile differs largely among 
five nights of observations. ($bottom$) Folded light-curve of the 
differential $r'$-band magnitudes with a best-fit period of 1.56278 hr 
(5626 s) proposed by \cite{rom12}. The phase zero is
 defined as MJD 56010.76808.  Note that both the amplitude of 
the modulation and the peak intensity changed by $\gtrsim$ 1 mag
and $\sim$  0.5 mag respectively, over the five nights of observations. 
}
\end{center}
\end{figure}

We made further deep follow-up observations of the field of 
CXOU J131145.71-343030.5, centered at (RA, DEC) = (197.940400\deg, 
-34.508306\deg), with the 105 cm Ritchey-Chr\'etien  
telescope at the Ishigakijima 
Astronomical Observatory in Japan. These observations were obtained 
on 2012 May 25 and started at 12:31:11:13 and ended 15:41:11:63 (UT).
The telescope in IAO is equipped with a tricolor camera that 
performs simultaneous imaging in the SDSS-$g'$ (hereafter, $g'$), 
$Rc$ and $Ic$ bands. 
The total net exposure amounts to 8400 s (300 s $\times$ 28 frames;
see Table 1 for the observation log).
All images were flat field and bias corrected. 
The absolute magnitudes were calibrated against the four 
reference stars shown in Figure 1.
In the $Rc$ and $Ic$ bands, the reference star magnitudes were based on the NOMAD catalog \citep{mon03b}. 
Because the target area was not covered by the SDSS, 
we employed a system conversion formula
of USNO-B $Bc$ and $Rc$ magnitudes \citep{ses06} for the four reference stars 
to obtain $g'$ magnitudes.
Moreover, the observed magnitudes were corrected for 
Galactic extinction using $A_{g'} = 
0.247$ mag, $A_{Rc} = 0.167$ mag, and $A_{Ic} = 0.120$ mag.

\begin{figure}
\begin{center}
\includegraphics[angle=0,scale=0.25]{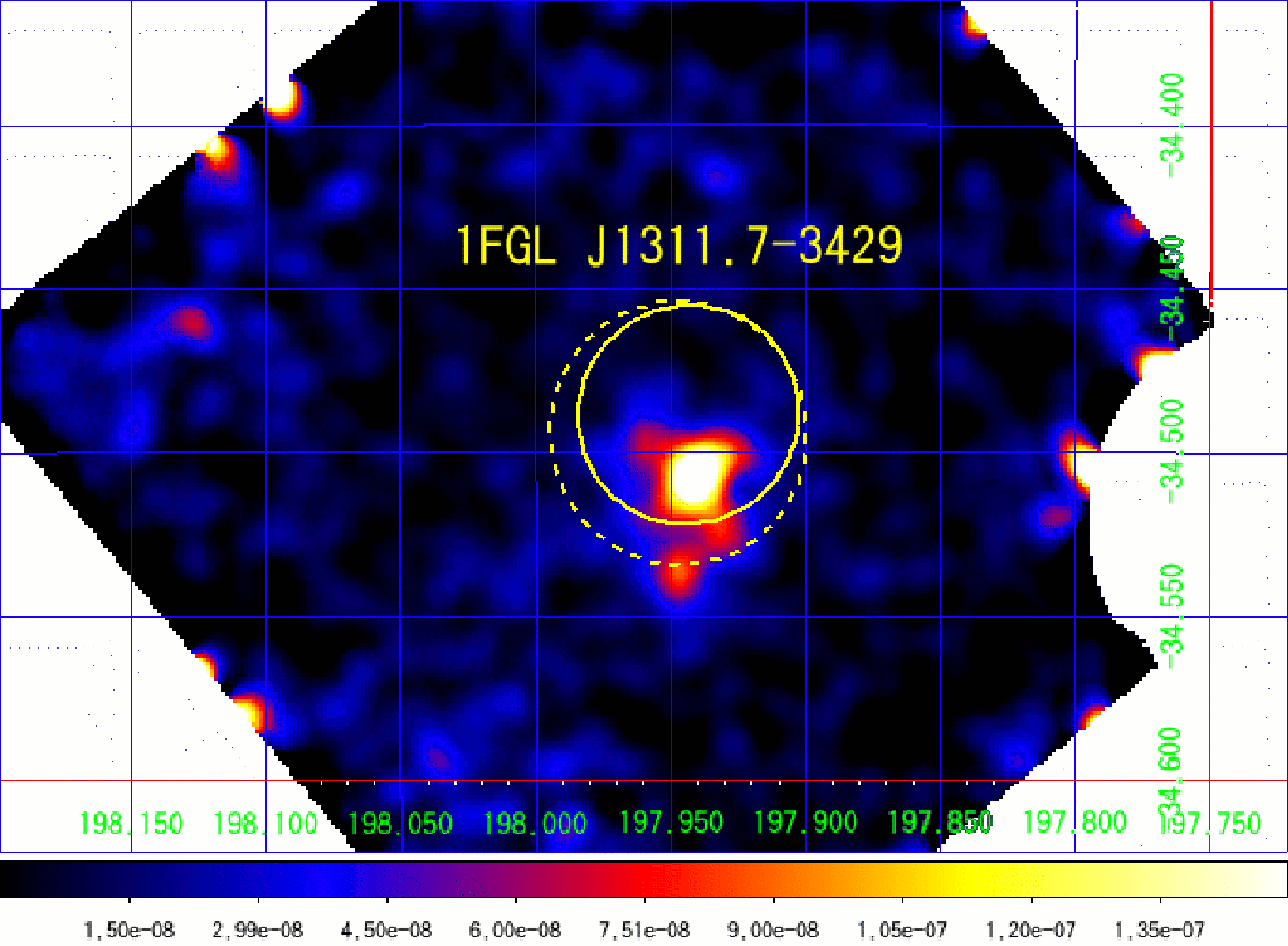}
\includegraphics[angle=0,scale=0.25]{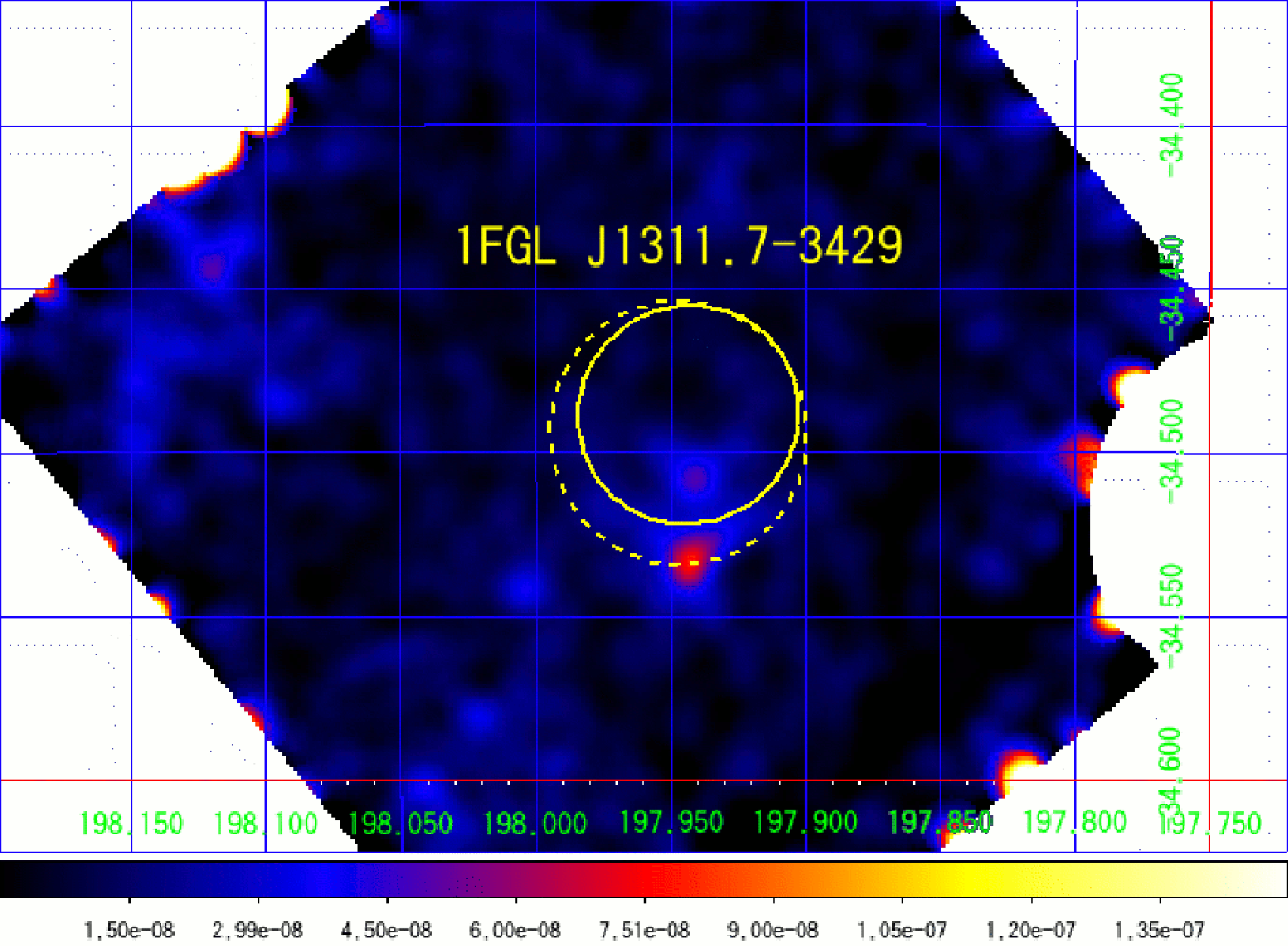}
\caption{$Suzaku$ XIS FI(XIS0+3) images of the 1FGL J1311.7-3429 region
 in the 0.5$-$10 keV photon energy range using the 
data during the first 20 ks ($top$; flare) and the last 74 ks
 ($bottom$; after the flare) in the 2009 observation obtained in AO4. 
The image shows the relative excess of
 smoothed photon counts (arbitrary units indicated in the color bar) and
 is displayed with linear scaling.  The thick solid ellipse denotes the
 95 $\%$ position error of 1FGL J1311.7-3429 in the 2FGL catalog, while
 the dashed ellipse shows that reported in the 1FGL catalog. Note the
 faint source to the south presented (and denoted as src B) in \citet{mae11} 
 is outside the 2FGL error circle.}
\end{center}
\end{figure}

We also conducted optical SDSS-$r'$ (hereafter $r'$) 
band monitoring observations with the Lulin One-meter Telescope 
\citep[LOT; ][]{lot}. Observations were conducted on 
five nights between 2012 March and May (see Table 1).
Photometric images with 300 s exposure  
were obtained using the PI1300B CCD camera. We performed the
dark-subtraction and flat-fielding correction using the appropriate
calibration data. 
For these LOT data, the four reference stars used in the IAO photometry
were saturated in the detector so could not be used.   
Instead, the LOT photometric results are presented as
differential magnitudes against two other fainter reference
stars (boxed in Figure 1).

\subsection{$Suzaku$ XIS}

As noted above, the first X-ray follow-up observation of 1FGL
J1311.7-3429 was conducted in 2009 as a part of AO-4 $Suzaku$ program
(PI: J.~Kataoka; OBS\_ID 804018010) 
aimed at observing an initial four out of 11 
unidentified \F objects at high Galactic latitude, 
$| b |$ $>$ 10$^{\circ}$ \citep[Maeda et al. 2011; see also][]{tak12}.
To further investigate the nature of the detected variable X-ray counterpart, 
we conducted the second $Suzaku$ observation of 1FGL J1311.7-3429 
in  2011 as a part of AO-6 program (PI: J.~Kataoka, OBS\_ID 706001010). 
The observation started at 2011 Aug 01 16:48:20 and ended at 
Aug 03 17:40:15. The total exposure amounted to 65.2 ks, and is 
almost twice as long as that obtained in AO-4 (see Table~1).
For both the AO-4 and AO-6 data analysis, 
we excluded the data collected during the time and up to 60 s after the 
South Atlantic Anomaly (SAA), and excluded data corresponding 
to less than 5$^{\circ}$ of the angle between Earth's limb 
and the pointing direction. Moreover, we excluded time windows during
which the spacecraft was passing through the low cut-off rigidity (COR) 
of below 6 GV.   We set the same source region to within a 1' 
radii around the respective X-ray flux maximum 
and the selection criteria for the data analysis for the two datasets were completely the same. 
Although $Suzaku$ also carries a hard X-ray detector (HXD), consisting of 
the PIN and GSO, hereafter, we do not use the data because the source 
is too faint to be detected with HXD/PIN or GSO.

\begin{figure}
\begin{center}
\includegraphics[angle=0,scale=0.47]{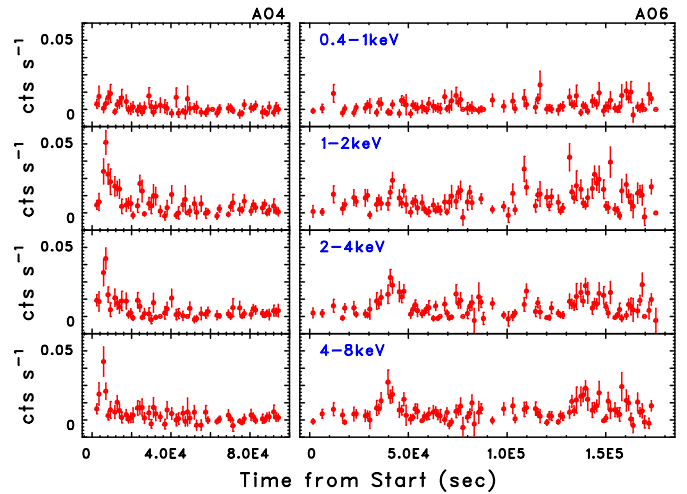}
\caption{Multi-band $Suzaku$ X-ray light curves of 1FGL J1311.7-3429
 obtained during AO4 (2009) and AO6 (2011) observations;
 0.4$-$1 keV, 1$-$2 keV, 2$-$4 keV, and 4$-$8 keV (from $top$ to
 $bottom$). The XIS-0, 1, 3 data are summed.}
\end{center}
\end{figure}

\section{Results}

\subsection{Optical/UV}

Figure 1 shows a multicolor image of 1FGL J1311.7-3429 
constructed from the IAO $g'$ ($blue$), $Rc$ ($green$), and $Ic$ ($red$) 
data from 2012 May 24.
The optical counterpart of the X-ray source is clearly detected 
in the image, but apparently the source is rather "blue" compared 
to reference stars and
nearby galaxies. Aperture photometry yielded average magnitudes 
of the source, $g'$ = 20.97 $\pm$ 0.13, $Rc$ = 21.17 $\pm$ 0.16, and $Ic$ $<$
18.38 (3 $\sigma$ upper limit).  
We also checked the temporal profile of the optical emission for 
the IAO data (Figure 2).  Relative photometry for the IAO images 
against the four field stars shows large amplitude 
($\Delta m$ $\sim$ 2 mag) quasi-sinusoidal modulation in the $g'$ and
$Rc$ bands
with a a timescale of 1.5 hr, as suggested by \cite{rom12}. 
%Figure 2 show the $g'$ and $Rc$-band light curves 
%obtained with the IAO on 2012 May 24.
%While the source is firmly detected in the $g'$ and $Rc$ bands, 
%we conservatively estimate 3 $\sigma$ upper limits for the $Ic$-band.

\begin{figure}
\begin{center}
\includegraphics[angle=0,scale=0.5]{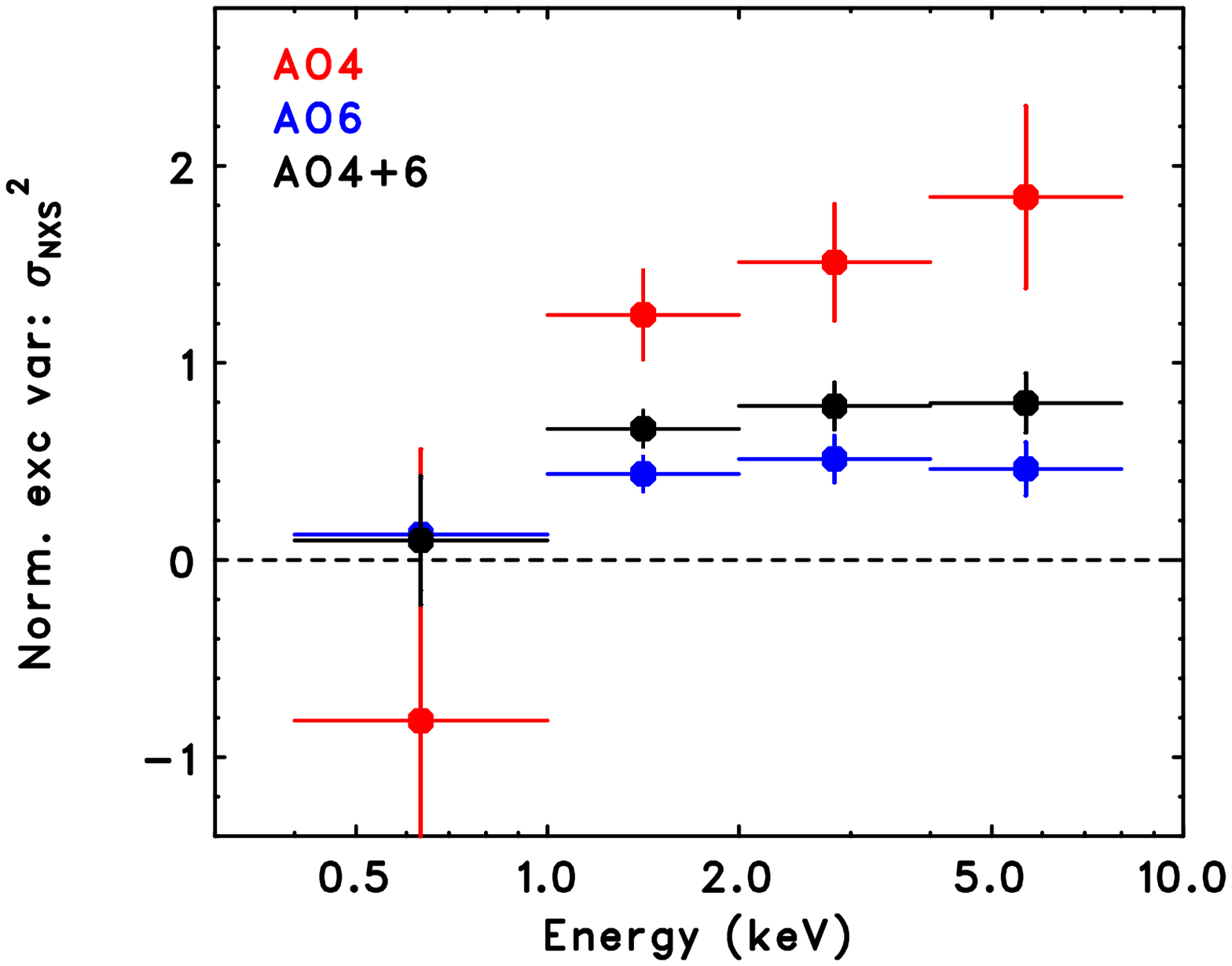}
\includegraphics[angle=0,scale=0.5]{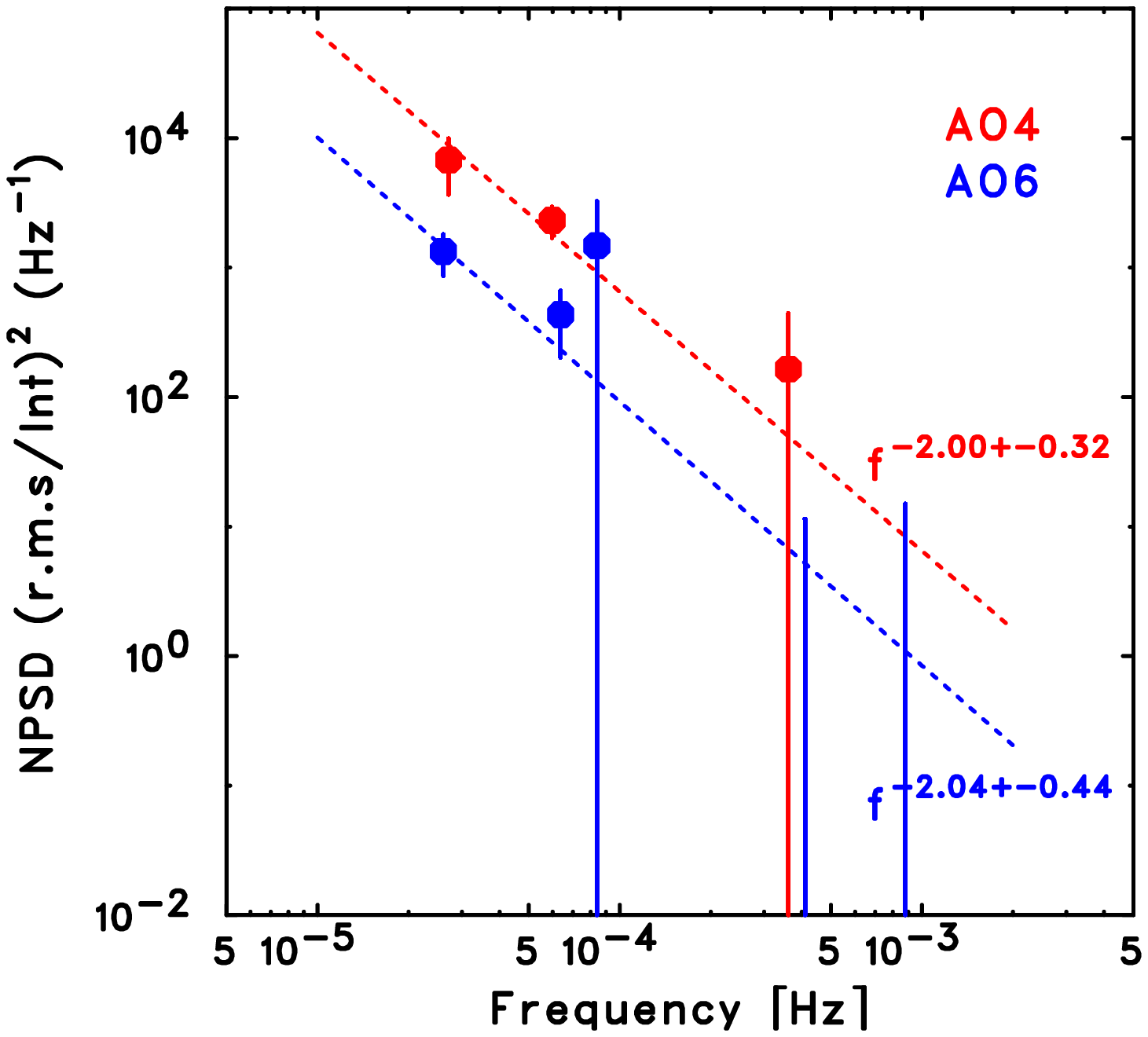}
\caption{($top$) Energy dependence of X-ray variability of 1FGL
 J1311.7-3429. The variability parameter, excess variance was calculated
 for the total exposures in AO4 (2009; $red$) and AO6 (2011; $blue$) 
in four energy bands (see Figure~5).
($bottom$) Normalized PSD (NPSD) calculated from the X-ray light curves
 of 1FGL J1311.7-3429. The dotted line shows the best-fitting power-law 
function of $\propto$$f^{-2.0}$.}
\end{center}
\end{figure}

To search for possible long-term variability, Figure 3 ($top$) compares 
the $r'$-band light curve reconstructed only from LOT data between 
2012 March and May. 
The differential magnitudes are derived by comparison with  
two reference stars in the field (see, Figure 1). 
Note, the $r'$-band magnitude of the reference stars remain 
constant within $\Delta$$m$ = 0.068 mag over 
the five nights of observations. 
Interestingly, the amplitude of modulation and 
the lightcurve profile seems to have changed among the five 
nights of observations. Specifically, a clear modulation of 
$\Delta m$ $\sim$ 2 mag is visible in 
the May 24 data, whilst $\Delta m$ $\sim$ 1 mag in the March 24 data, and 
almost unseen ($\Delta m$ $\lesssim$ 0.2 mag) in the March 26 data. 
Moreover, the peak magnitude differs by $\Delta m$ $\sim$ 0.5 mag 
among the five nights, much larger than the fluctuations in the
magnitudes of  reference stars. Figure 3 ($bottom$) shows the 
folded light-curves of differential $r'$-band magnitudes with 
a best-fit period of 1.56278 hr (5626 s) proposed by \cite{rom12}.
Phase zero is defined here as MJD 56010.76808, so that 
the time of the observed minimum $r'$-band magnitude in the May 24 data 
is set at orbital phase $\phi$ = 1.0.
The peak $r'$-band magnitude is around 
$\simeq$20.5 for five nights of data from 2012 March and April.
Note, this is exactly consistent with what has been observed with 
IAO on May 25 (Figure 2; $right$). 
To further investigate the temporal variability/flaring in the UV data, 
we also re-analyzed the archival $Swift$ UVOT data. However, the 
UVOT exposures for each filter ($b$, $u$, $uvw1$, $uvm2$, $uvw2$) 
were too short to search for variability (see Table 1).

\subsection{$Suzaku$ XIS}

During the AO-4 observation,
significant X-ray variability was detected at the beginning 
of the observation, where the count rate changing by a factor of 
10 \citep{mae11}. Figure 4 compares the $Suzaku$ XIS FI (XIS0+3) 
images in the 0.5$-$10 keV range, reconstructed by using the 
data during the first 20 ks ($top$) and the last 74 ks ($bottom$). 
The images are corrected for exposure and vignetting, and are non-X-ray 
(detector) background subtracted. The images were in addition  
smoothed by a Gaussian function with $\sigma$ = 0.'17, following the 
procedure given in \cite{mae11}. Note that the source is clearly 
detected in the first 20 ks, but almost unseen in the last 74 ks.
Also \cite{che12} argues that the 0.3$-$10 keV $Swift$ XRT count rate of 
(6.1$\pm$1.8)$\times$10$^{-3}$ counts s$^{-1}$ is equivalent 
to a 0.5$-$8 keV flux of $\simeq$ 3.1$\times$10$^{-13}$ ergs cm$^{-2}$ 
s$^{-1}$, indicating a $\sim$3$\times$ brighter source about one year 
prior to the $Chandra$ observation, and providing further evidence 
of long-term X-ray variability in this source. 

\begin{figure}
\begin{center}
\includegraphics[angle=0,scale=0.5]{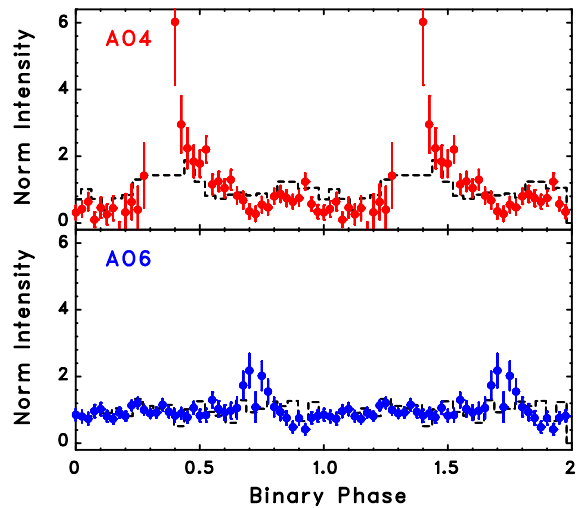}
\caption{The folded X-ray light curves of 1FGL J1311.7-3429 
reconstructed from the XIS data taken in $Suzaku$ AO4 ($top$; $red$)
 and AO6 ($bottom$) observations. All the XIS0, 1, 3 data are summed
 in the energy range of 0.4$-$10 keV. The phase zero is
 defined as MJD 56010.76808. The vertical axis 
represents the source counts divided by the average source intensity
in the frame. $Dashed$ $line$ represents 
the folded X-ray light curves of a nearby field source 
\citep[src B in][]{mae11}
to confirm that
 observed excess is not due to artifacts caused by orbital gaps in the 
$Suzaku$ data.}
\end{center}
\end{figure}

\begin{deluxetable}{lcc}
%\tabletypesize{\scriptsize}
%\rotate
\tablecaption{Fitting Parameters of $Suzaku$ data for the Power-law
 Model}
\tablewidth{0pt}
\tablehead{
\colhead{Parameter} &  \colhead{AO4} & \colhead{AO6} \\
\colhead{} &  \colhead{Value and Errors$^a$} & \colhead{Value and Errors$^a$}}
\startdata
$N_{\rm H}$ (10$^{20}$ cm$^{-2}$)  & 0.33$^{+0.15}_{-0.13}$ & $<$ 0.07  \\
$\Gamma$  & 1.71$^{+0.18}_{-0.17}$  &  1.26$^{+0.08}_{-0.07}$ \\
Flux (0.5-2 keV)$^b$  & 0.99$^{+0.27}_{-0.19}$  & 0.83$^{+0.08}_{-0.06}$  \\
Flux (2-10 keV)$^b$  & 1.78$^{+0.16}_{-0.15}$  & 2.96$\pm$0.15  \\
\tableline
$\chi^2$ (dof)  & 17.4 (15)  & 17.6 (15) \\
$P$($\chi^2$)  & 0.29 & 0.29  \\ 
\tableline
\enddata 
\tablecomments{$^a$: all errors are 1$\sigma$}
\tablecomments{$^b$: in unit of 10$^{-13}$ erg cm$^{-2}$ s$^{-1}$}
\end{deluxetable}

Figure 5 summarizes the X-ray light curve of 1FGL J1311.7-3429 
thus obtained during the AO-6 observation and compared with those obtained 
in the AO-4 observation.  All the XIS0, 1, 3 count rates are summed in 
the various energy bands of 0.4$-$1 keV,  1$-$2 keV, 2$-$4 keV and 4$-$8 keV 
(from $top$ to the $bottom$). Note that X-ray variability is clearly 
seen above 1 keV, but not in the 0.4$-$1 keV light curve.  
To see this more quantitatively, we calculated  the normalized 
excess variance $\sigma_{NXS}^2$ 
\citep[Vaughan et al. 2003; see also][]{kat07} 
for the AO4, AO6 light curves and for the combined AO4+AO6 
light curves. The $\sigma_{NXS}^2$ parameter 
is an estimator of the intrinsic source variance after subtracting 
the contribution expected from measurement errors. As shown in 
Figure 6 ($top$), $\sigma_{NXS}^2$ is consistent with zero for 
the $0.4-1$ keV band, whilst almost constant (within error bars) and 
significantly positive above 1 keV.\footnote{The fractional root mean 
square variability amplitude, $F_{\rm var}$ is often chosen in
preference to $\sigma_{NXS}^2$, although the two convey exactly the
same information. Here, we prefer $\sigma_{NXS}^2$ simply because 
variability is not significant for the $0.4-1$ keV light curve and negative 
values of $\sigma_{NXS}^2$ are possible.}   

Figure 6 ($bottom$) presents the normalized power spectrum density 
\citep[NPSD; ][]{hay98} which is a technique for calculating the PSD of 
unevenly sampled light curves. Note that data gaps are unavoidable for 
low Earth orbit X-ray satellites like $Suzaku$. Since the orbital period of 
$Suzaku$ is $\sim$ 5760 s, Earth occultations create artificial periodic gaps 
every 5760 s in the data, even when the source is continuously monitored.
To calculate the NPSD of our data sets, we made light curves of 
two different bin sizes of 256 s and 5760 s for each of the AO4 and 
AO6 light curves 
\citep{kat01}.  The NPSD calculated for each light curves is 
well represented by a steep power-law with $P(f)$ $\propto$
$f^{-2.00\pm0.32}$ (AO4) and $P(f)$ $\propto$ $f^{-2.04\pm0.44}$ (AO6).
Note that the amplitude of the NPSD is larger for the AO4 ($red$) data than 
in the AO6 ($blue$) data. This is due to large flare observed in the
first 20 ks of the AO4 observation (Figure 4) and is consistent 
with the larger 
value of  $\sigma_{NXS}^2$ compared to variability during AO6 observation 
(Figure 6 $top$).

\begin{figure}
\begin{center}
\includegraphics[angle=0,scale=0.5]{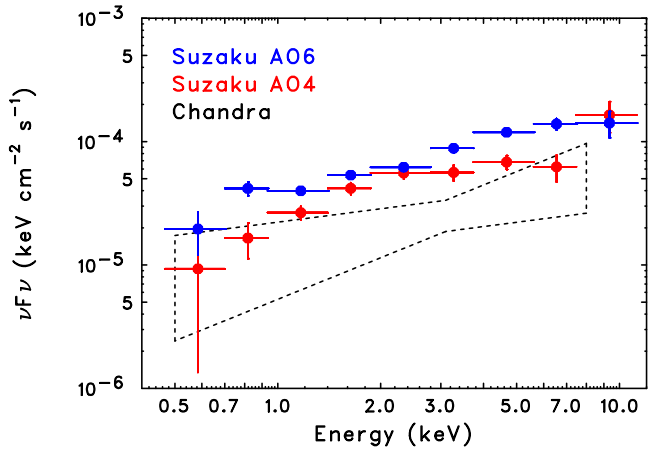}
\caption{Comparison of the average X-ray spectra of 1FGL J1311.7-3429
 observed in 2009 (AO4; $red$) and 2011 (AO6; $blue$) $Suzaku$
 observations. The X-ray data points represent the weighted mean of
 XIS0, 1 and 3. In addition, a bow-tie shows the best-fit parameters of the 
corresponding $Chandra$ source, CXOU J131145.7-343030, observed in 2010
\citep{che12}.}
\end{center}
\end{figure}

Figure 7 shows the folded X-ray light curves from the $Suzaku$ AO4
($top$) and AO6 ($bottom$) data sets. The XIS0,1 and 3 data are all 
summed in 0.4$-$10 keV band. 
The phase zero is defined as MJD 56010.76808 as in 
the optical folded light curves shown in Figure 3 ($bottom$).  
The vertical axis 
represents the source counts divided by the average source intensity 
in the frame. Although there is a regular exposure gap 
for phases $\phi$ = 0.3$-$0.4 due to Earth occultations, 
there appears to be a clear excess at phases $\phi$ = 0.4$-$0.6 in AO4 data 
(see, also \cite{rom12}). Moreover, we found similar 
structure in the AO6 data, but at different phases of $\phi$ = 0.6$-$0.8 
with much smaller amount of excess. To check that this excess is $not$ 
due to artifacts caused by periodic gaps due to Earth occultation, 
we also made the folded light curves of the nearby X-ray source 
in the same field of view (src B in \cite{mae11}, or CXOU
J131147.0-343205 in \cite{che12}; see Figure~4), whose X-ray flux is
comparable to the average X-ray flux of 1FGL J1311.7-3429. The results 
presented as dashed lines in Figure 7 
show no clear excess for the nearby source. This suggests 
that the X-ray excesses in the folded lighturves are most likely due to 
orbital motion of 1FGL J1311.7-3429 itself.

Finally, Figure 8 compares the unfolded X-ray spectrum for 1FGL
J1311.7-3429 averaged over the AO4 ($red$) and AO6 ($blue$) observations. 
The best model fits for both observations consist of power-law 
continua with photon indices, $\Gamma$ $\simeq$ 1.3$-$1.7. 
Significant curvature (or deficit of photons below 2 keV) was 
observed only in the 2009 data, which was tentatively modeled by 
an excess value of Galactic column density $N_{\rm H}$; $N_{\rm H}$ 
is however consistent with zero for the 2011 data. 
The model fitting results are summarized in Table 2. 
Although the nearby X-ray source (src B in \cite{mae11}, or CXOU
J131147.0-343205 in \cite{che12}; see Figure~4) is located only 1.6' apart 
to the south, contamination from this source and other nearby 
faint sources is estimated to be less than 5 $\%$ 
for the applied region of interest of 1'.

\section{Discussion and Conclusions}

During the first (AO4; 2009) and second (AO6; 2011) $Suzaku$ 
observations, we detected significant variability in 1FGL J1311.7-3429 characterized 
by repeated flaring activity, with a timescale of $\sim$ 10 ks.
The variability is only clearly seen above 1 keV. 
The NPSD is well characterized by $P(f)$ $\propto$ $f^{-2}$, as is 
the case for the X-ray variability of various classes of AGN 
including Seyferts and blazars \citep[e.g.,][]{hay98, vau03}. 
The latter class constitutes the majority of \F sources 
but the X-ray variability timescales of blazars are in general 
somewhat longer,  typically $\simeq$ 1 day \citep[e.g.,][]{kat01}. 
Such ``red-noise'' PSD behavior is also 
observed in X-ray (Galactic
black hole and neutron star) binary systems, but on much shorter time scales. 
For example, variability as short as $\sim$ 1$-$10 ms has been observed 
for the famous Galactic black hole source Cyg X-1 
\citep[e.g.,][]{mee84, hay98}.

\begin{figure}
\begin{center}
\includegraphics[angle=0,scale=0.47]{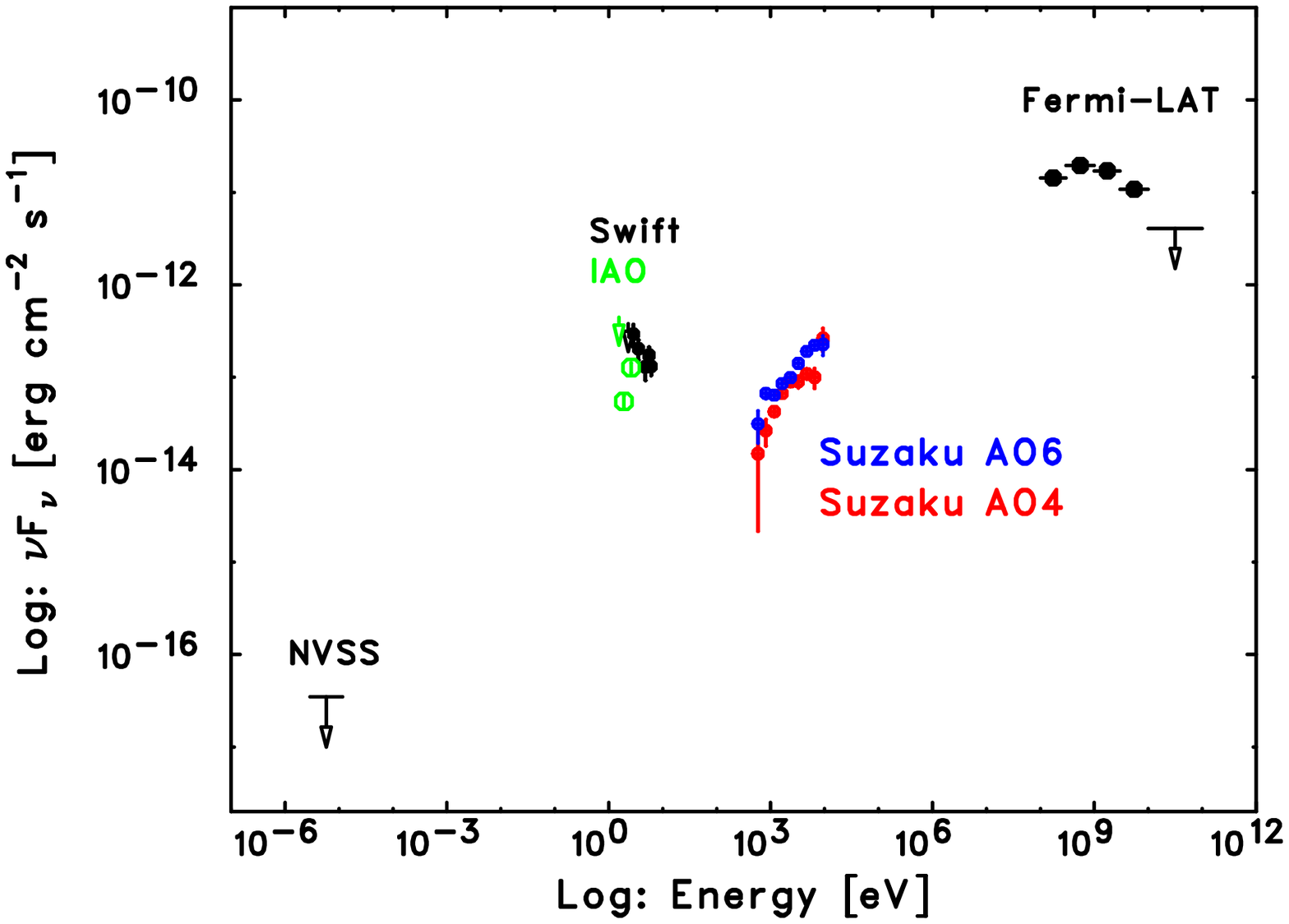}
\includegraphics[angle=0,scale=0.47]{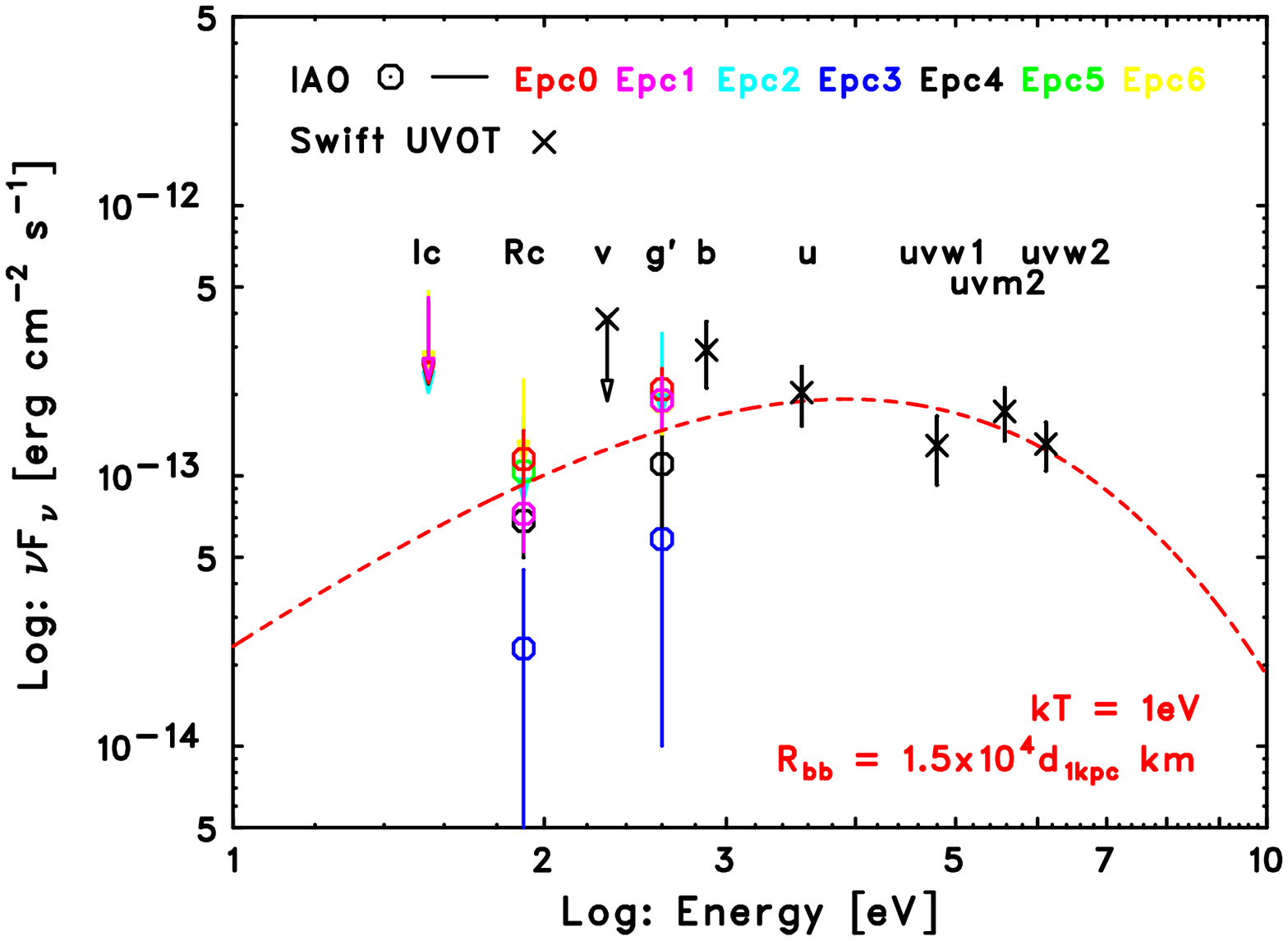}
\caption{($top$) Broadband spectrum of 1FGL J1311.703429. The X-ray data are
 as presented in Figure 8. The gamma-ray data points are
 taken from the 2FGL catalog \citep{2FGL}. The radio upper limit of 2.5 mJy at 1.4 GHz is taken from the NVSS catalog \citep{NVSS}.
 The optical and UV data represents
 $Swift$ UVOT data (see Maeda et al. 2011 and Cheung et al. 2012).
($bottom$) Close-up of the broadband spectrum in the optical-UV range.
Epochs 0$-$6 (1200 s each) correspond to time ranges 
defined in Figure 2.
Note the optical-UV spectrum is well fitted by a blackbody model 
of $kT$ $\sim$ 1eV with a radius of the emission  volume 
of $R_{bb}$ $\simeq$ 1.5$\times$10$^4$ $d_{\rm kpc}$ km, where $d_{\rm
 kpc}$ is distance to the source in unit of 1 kpc.}
\end{center}
\end{figure}

Significant variability has also been observed in the optical ($g'$, $Rc$,
$r'$), where it is rather a quasi-sinusoidal flux modulation 
with a 1.56 hr period, as recently 
reported by \citet{rom12}.
Moreover, we also found that the modulation profile, including 
the amplitude of modulation and peak intensity, has 
changed largely among the six nights of observations.  
The apparent modulation of magnitude observed in both the IAO and LOT 
data is quite similar to those observed in 1FGL J2339.7-0531, 
which is characterized by a 4.63-hr 
orbital period in optical and X-ray data \citep{rom11,kon12}.
Note that 1FGL J2339.7-0531 is now suggested to be a ``radio-quiet'' 
gamma-ray emitting black-widow MSP with a $\simeq$0.1 $M_{\odot}$ late-type 
companion star, viewed at inclination $i$ $\simeq$ 57$^{\circ}$. 
Moreover, this compact object companion in 1FGL J2339.7-0531 is strongly 
heated, with $T_{\rm eff}$ varying from $\sim$ 6900 K (superior 
conjunction) to $<$ 3000 K  at minimum \citep{rom11}. 

\begin{figure}
\begin{center}
\includegraphics[angle=0,scale=0.53]{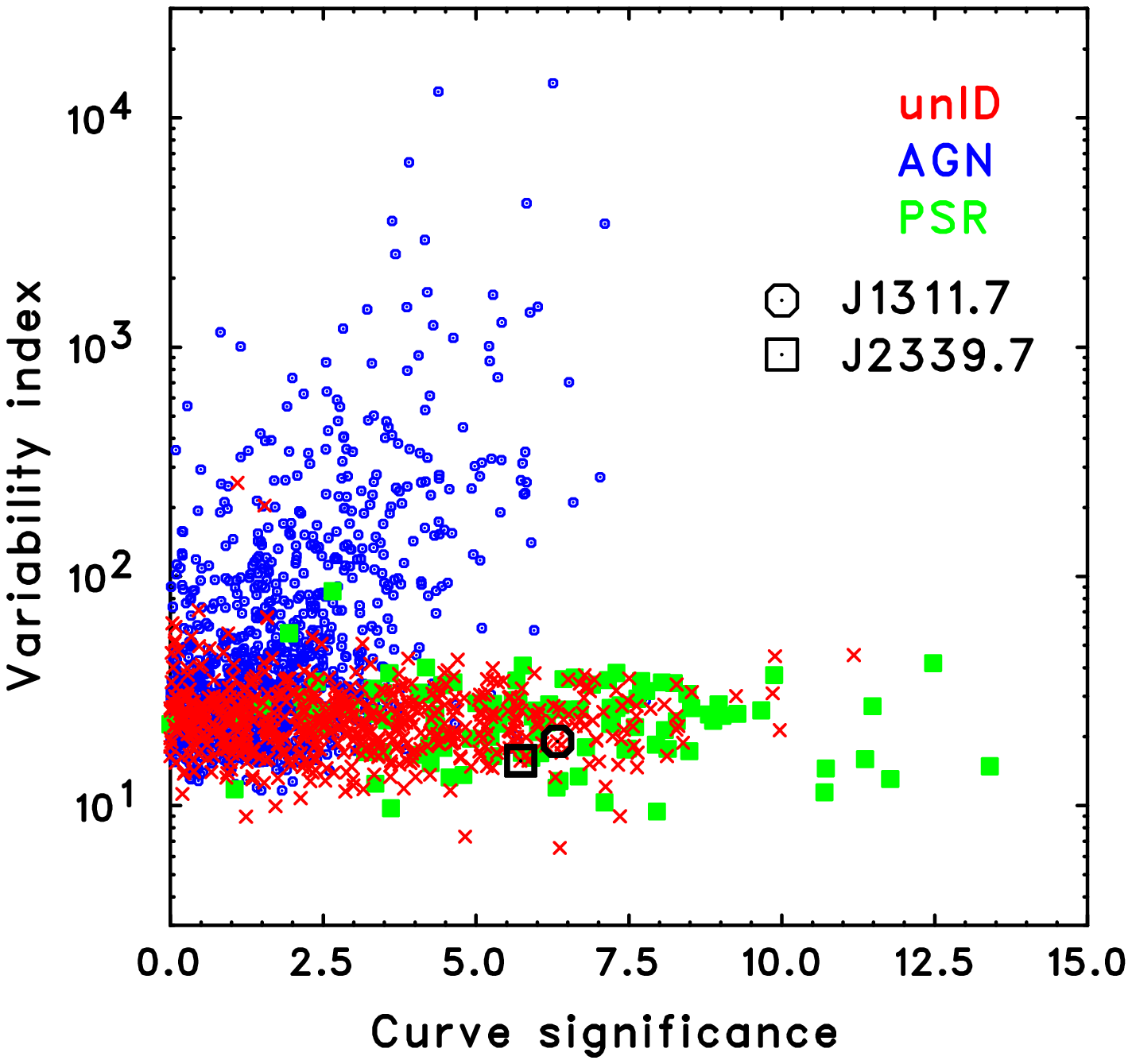}
\includegraphics[angle=0,scale=0.53]{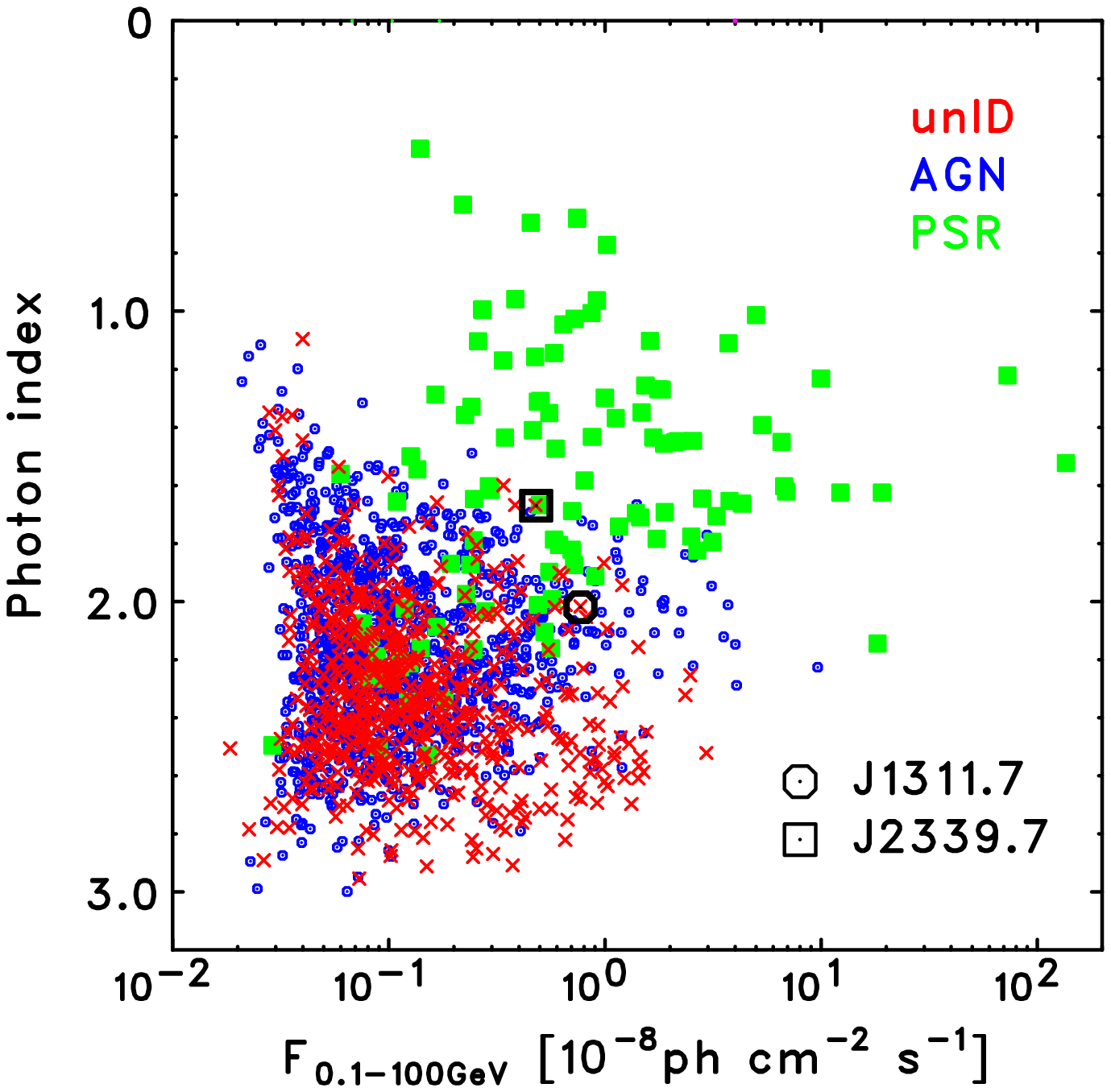}
\caption{($top$) Comparison of the 2FGL Variability Index versus
 Curvature Significance for associated sources (AGN: $blue$, PSR: $green$)  
 and unassociated sources ($red$). ($bottom$) Comparison of the
 2FGL Photon Index versus $0.1-100$ GeV Flux for the associated sources 
and unassociated sources. In both 
panels, a separation between the AGN and PSR
populations is evident. Note that 1FGL J1311.7-3429 is situated in the
typical PSR region in the $top$ panel, whilst at the boundary of
 AGN/PSR sources in the $bottom$ panel. 
}
\end{center}
\end{figure}

In this context, the spectral energy distribution of 1FGL J1311.7-3419 
 from radio to gamma-rays (Figure 9, $top$) may provide 
some hints on the nature of this mysterious source.  1FGL J1311.7-3419
 has been the subject of both radio pulsar counterpart searches and 
blind searches for gamma-ray pulsations, but to date no pulsed emission 
has been detected \citep{ran11}, and the situation is quite similar to the case for 
1FGL J2339.7-0531. There are no NVSS radio sources catalogued within the 2FGL 
ellipse of 1FGL J1311.7-3429 down to 1.4 GHz flux completeness limit 
of $\simeq$2.5 mJy \citep{NVSS}, which is given as an arrow in Figure 9
 ($top$).
From the 0.1$-$100 GeV gamma-ray flux listed in the 2FGL catalog,  
$F_{\gamma}$ $\equiv$ $F_{\rm 0.1-100GeV}$ $\simeq$
 6.2$\times$10$^{-11}$ erg cm$^{-2}$ s$^{-1}$,  we obtain 
$F_{\gamma}$/$F_X$ $\simeq$ 300, $F_{\gamma}$/$F_R$ 
$\ge$ 1.7$\times$10$^6$ for 1FGL J1311.7-3429, where $F_X$ and 
$F_R$ are the X-ray and radio fluxes measured in 2$-$10 keV and at 
1.4 GHz, respectively. Note, these values are almost 
comparable to those measured in 1FGL J2339.7-0531,  with 
$F_\gamma$ $\simeq$ 3.0$\times$10$^{-11}$ erg cm$^2$
 s$^{-1}$, $F_{\gamma}$/$F_X$ $\simeq$ 150 and $F_{\gamma}$/$F_R$ 
$\ge$ 8.7$\times$10$^5$. Although its flat spectrum X-ray continua
and relatively high $\gamma$-ray-to-X-ray energy flux ratio 
(of $\gtrsim$ 100) is typical of the sub-class of blazars 
\citep[FSRQ;][]{2FGL}, the non-detection in the radio as well as 
its curved gamma-ray spectrum seems to strongly disfavor an AGN association
 for 1FGL J1311.7-3429. 
   
To further support the MSP association of 1FGL J1311.7-3429,
Figure 9 ($bottom$) shows a close-up of the SED of 
1FGL J1311.7-3429 in the optical/UV bands, reconstructed from 
IAO and $Swift$ UVOT data. Observed magnitudes were converted to flux 
density based on \cite{fuk95}. 
$Red$ $dashed$ $line$ indicates a
 tentative fit with a blackbody model of $T$ $\simeq$ 1.2$\times$10$^4$ K 
(or $kT$ $\simeq$ 1 eV) assuming an emission volume radius of 
$R_{bb}$ $\simeq$ 1.5$\times$10$^4$ $d_{\rm kpc}$ km, 
where $d_{\rm kpc}$ is the distance to the source in units of 1 kpc. 
Therefore, the optical/UV spectrum of 1FGL J1311.7-3429 seems compatible 
with what is expected from a companion star of a radio-quiet MSP 
like 1FGL J2339.7-0531.  
A slightly higher temperature than 1FGL J2339.7-0531 
\citep[$kT$ $\simeq$ 0.3$-$0.6 keV;][]{rom11} 
may indicate $\times$2 smaller orbital radius, $\sim$
5$\times$10$^5$ km, for the 1FGL J1311.7-3429 binary system, 
assuming that the mass of the companion star is 0.1 $M_{\odot}$ and 
the pulsar spin-down luminosity is  $L$ $\simeq$ 10$^{34}$ erg s$^{-1}$ 
\citep[parameters suggested for 1FGL J2339.7-0531 binary system;][]{rom11}.
A change of the modulation profile observed in the $r'$-band 
may be accounted for by rapid changes in the companion star temperature, 
but this remains uncertain.

Since both the optical and X-ray light curves in the case of
1FGL J2339.7-0531 clearly exhibits a 4.63-hr orbital modulation 
\citep{rom11,kon12}, the detection of periodicity in the X-ray 
light curve of 1FGL J1311.7-3429 is also likely. 
In fact, the folded X-ray light curve exhibits an  
excess feature around $\phi$ = 0.5. As indicated by \cite{rom12}, 
this is presumably pulsar superior conjunction, although both the 
normalized intensity as well as the phase peak 
appear to have changed substantially between the 2009 (AO4) and 2011 (AO6) 
observations. We therefore expect that 
X-ray variability consists of at least two different components -- 
one associated with the binary motion as for the optical data, 
and the other is rather random fluctuation well represented by 
the NPSD of $P(f)$ $\propto$ $f^{-2}$, whose physical origin is 
still unknown but possibly related with perturbations associated with
shock acceleration.

Such flaring X-ray variability has not yet  
been observed for 1FGL J2339.7-3429, but solely in 1FGL J1311.7-3429.
In a review of X-ray emission from MSPs, \cite{zav07} describes 
three primary sources of X-ray emission: (1) intra-binary shock, 
(2) the neutron star (NS) itself, and (3) pulsar wind nebula outside 
the binary system \citep[see, also][]{arc10}. In fact,  
thermal emission of $kT$ $\sim$0.1$-$0.2 keV is often 
observed from MSPs, which is thought to arise from the surface 
of the NS and is steady with time \citep[e.g.,][]{mar11,mae11}. 
More recently, 
thermal emission of $kT$ $\simeq$ 0.1 keV was also detected from 
1FGL J2339.7-0531 (Kong et al. 2012, in prep), which is again steady 
despite large flux modulation associated with binary motion being observed
above 2 keV.  Therefore, the fact that X-ray 
variability is not clearly seen below 1 keV may suggest there could 
be some contribution from the surface of the NS, although from 
the spectral fitting, it is not statistically significant. 
Then the variable, hard X-ray emission could 
arise from the intra-binary shock rather than the nebula because variability 
as short as $\simeq$ 10 ks is unlikely to originate from the extended 
pulsar wind nebula\footnote{Variability has been observed 
in some pulsar wind nebula in X-rays and gamma-rays, but with typical 
timescales of a week to months \citep[e.g.,][]{pav01,Crab}.}.  
Such a shock  could readily produce gamma-ray emission 
\citep{aro93}. If localized, it could easily account for the 
orbital modulation as seen in the X-ray emission from 
1FGL J2339.7-0531. But if material leaving the companion star, 
either through Roche-lobe overflow or a stellar wind
 is non-uniform or patchy, we might expect random flaring  
activity as we see in the X-ray data of 1FGL J1311.7-3429.
    
We can also speculate on the nature of 1FGL J1311.7-3429 based 
solely on the gamma-ray properties, an approach  
already applied to the 1FGL unIDs in \cite{UnIDstat}. Figure 10 ($top$) 
presents a comparison of the 2FGL associated (either AGN ($blue$) or 
PSRs ($green$)) and unassociated sources ($red$) in the
\textsc{Variability\_index} and \textsc{Signif\_curve} plane. 
Apparently, 1FGL J1311.7-3429 is situated in the typical PSR regions of this
diagnostic plane.
Similarly, Figure 10 ($bottom$) plots the distribution of PSRs and 
AGN in the \textsc{Photon index} versus \textsc{F$_{\rm 0.1-100 GeV}$} 
plane. In this case, 1FGL J1311.7-3429 is at the boundary of 
AGN and PSR sources, so is still consistent with a PSR association. 
For comparison, we also plot the gamma-ray parameters for 1FGL J2339.7-0531. 
In both panels, the gamma-ray properties of 1FGL J1311.7-3429 and 
1FGL J2339.7-0531 
are quite similar. Again, this supports the idea that 
1FGL J1311.7-3429 is a black-widow system and may be a second 
example of a ``radio-quiet'' MSP after 1FGL J2339.7-0531. 

Finally, if the rapid X-ray flaring variability observed with 
$Suzaku$ may be due 
to inhomogeneity of shock material and/or rapid changes in the
beaming factor, this could be expected to 
occur also in gamma-rays, as suggested by a smooth connection of 
the spectrum between X-ray and gamma-ray energies. Moreover, we may 
see correlated variability also in the optical, which may be 
related to the change of modulation profile in the  $r'$-band 
magnitudes we see in Figure 3. Unfortunately, such fast variability 
is difficult to observe in gamma-rays as we referred to one-month 
binned light curve in Section 1, despite the excellent sensitivity of 
$Fermi$-LAT. The low gamma-ray statistics also make 
it difficult to run a cross-correlation in order to measure 
any possible correlated optical, X-ray and gamma-ray variability.  
Further continuous investigation is necessary to confirm the origin of 
``variable'' X-ray emission observed in the 1FGL J1311.7-3429 system.

\acknowledgments 

This work is partially supported by the Japanese Society for the 
Promotion of Science (JSPS) KAKENHI 19047002. 
Work by C.C.C. at NRL is supported in part by NASA DPR S-15633-Y.
We would like to thank the anonymous referee for his/her helpful 
comments that improved the manuscript.

\end{document}